# Superconducting properties of La-substituted Bi-2201 crystals

Ya G Ponomarev[1], N Z Timergaleev[1], Kim Ki Uk[1], M A Lorenz[2], G Müller[2], H Piel[2], H Schmidt[2], A Krapf[3], T E Os'kina[4], Yu D Tretyakov[4] and V F Kozlovskii[4]

[1] M.V. Lomonosov Moscow State University, Faculty of Physics, 119899 Moscow, Russia
[2] Bergische Universität Wuppertal, Fachbereich Physik, Gaußstr. 20, D-42097 Wuppertal, Germany
[3] Humboldt-Universität zu Berlin, Institut für Physik, Invalidenstr. 110, D-10115 Berlin, Germany
[4] M.V. Lomonosov Moscow State University, Faculty of Chemistry, 119899 Moscow, Russia

**ABSTRACT:** The superconducting gap $\Delta_s$ has been measured in $Bi_2Sr_{2-x}La_xCuO_{6+\delta}$ single crystals in a wide range of temperatures $4.2\ K \leq T \leq T_c$ by point-contact and tunnelling spectroscopy for current in **c**-direction. The value of $\Delta_s(4.2\ K)$ was found to scale with the critical temperature $T_c$ in the whole range of doping levels with the ratio $2\Delta/kT_c = 12.5 \pm 2$. The closing of the gap $\Delta_s$ at $T = T_c$ has been registered in the underdoped, optimally doped as well as in the overdoped samples.

## 1   INTRODUCTION

Recently it has been proposed by Deutscher (1999) that for the underdoped copper oxide superconductors there exist two different gap energies $\Delta_p$ and $\Delta_s$. The larger gap (pseudogap) $\Delta_p$, measured by angle-resolved photoemission spectroscopy (ARPES) or tunnelling spectroscopy, is supposed to be a half of the energy required to split an incoherent Cooper pair. The smaller (superconducting) gap $\Delta_s$, determined by electron Raman scattering or Andreev spectroscopy, is associated with a superconducting state. According to Deutscher (1999) $\Delta_s$ scales with the critical temperature $T_c$ on doping, while $\Delta_p$ in underdoped samples continues to grow as $T_c \rightarrow 0$. It remains unclear why tunnelling spectroscopy should register in the underdoped HTSC samples only one excitation energy ($\Delta_p$). Probably one possible explanation comes from the STM tunnelling measurements (NIS contacts) on the underdoped $Bi_2Sr_2Ca_1Cu_2O_{8+\delta}$ (Bi-2212) single crystals (Renner et al 1998). The authors claim that a superconducting gap $\Delta_s$ does not depend on temperature and transforms directly into a pseudogap $\Delta_p$ of exactly the same magnitude at $T > T_c$. Unfortunately this is in conflict with the results obtained on SIS Bi-2212 contacts (Miyakawa et al 1998 and 1999) which give a convincing evidence that a superconducting gap $\Delta_s$ in the underdoped Bi-2212 crystals closes at $T = T_c$. The absence of scaling of $\Delta_s$ with $T_c$ on doping in the underdoped Bi-2212 crystals has been reported in several STM studies (Miyakawa et al 1998 and 1999 and Nakano et al 1998). In contrast Matsuda et al (1999) observed that $\Delta_s$ passes through a maximum at the optimal doping level in accordance with Deutscher (1999).

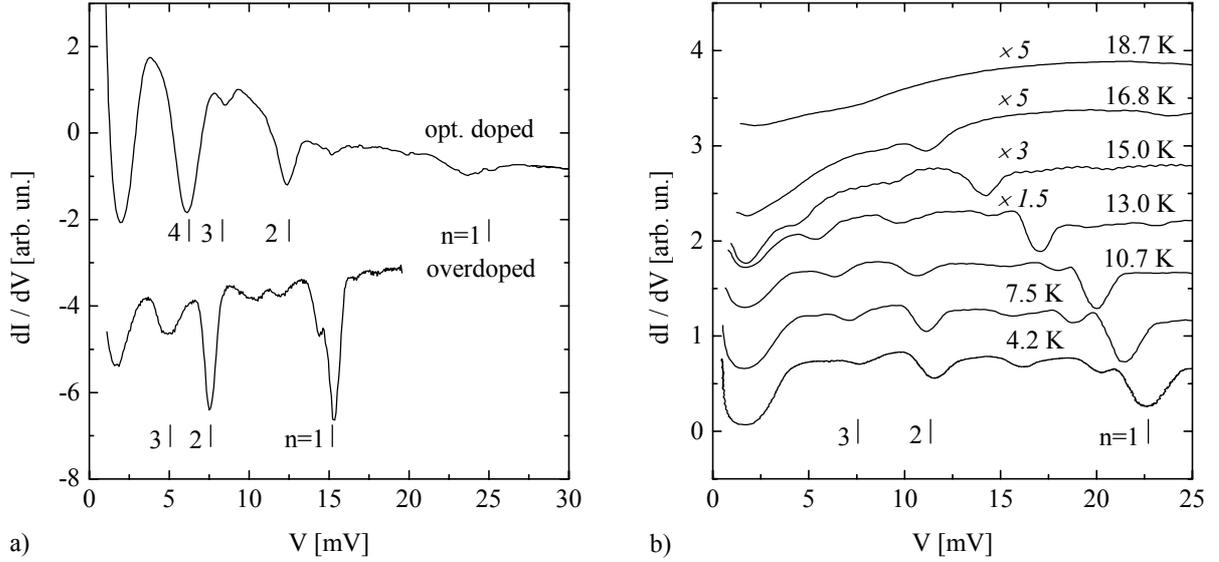

Fig.1. a) Subharmonic gap structure (SGS) in the $dI/dV$-characteristics of SNS contacts in optimally doped and overdoped Bi-2201(La) single crystals at $T = 4.2$ K with current in **c**-direction. The bias voltages $V_n = 2\Delta_s/en$ corresponding to the dips in the SGS are marked by bars.  b) Temperature dependence of the SGS in the $dI/dV$-characteristics of a SNS contact in an underdoped Bi-2201(La) single crystal with $T_c = 19.5$ K. Some of the characteristics are magnified by indicated factors.

It was interesting for us to verify whether the model proposed by Deutscher (1999) is applicable in the case of single-plane $Bi_2Sr_{2-x}La_xCuO_{6+\delta}$ crystals (Bi-2201(La)) with the value of the coherence length exceeding significantly that for other members of BSCCO-family (Yoshizaki et al 1994). In this system the entire range from overdoped to underdoped samples can be covered by substitution of Sr by La without damaging significantly the crystalline structure (Khasanova et al 1995 and Yang et al 1998). The existence of a pseudogap $\Delta_p$ has already been reported for Bi-2201(La) (Harris et al 1997 and Vedeneev 1998) but no detailed studies of the doping dependence of $\Delta_s$ have yet been done.

## 2     EXPERIMENTAL RESULTS AND DISCUSSION

The $Bi_2Sr_{2-x}La_xCuO_{6+\delta}$ crystals with the actual La concentration $0.1 \leq x \leq 0.5$ and the maximum $T_c$ of about 25 K were grown from copper-oxide-rich melt. The X-ray diffraction pattern of the Bi-2201 phase showed pseudo-tetragonal symmetry with **a** = 0.538 - 0.539 nm for all tested powders and crystals. Furthermore the **c**-axis lattice parameter demonstrated a strong dependence on La doping in accordance with Khasanova et al (1995) and Yang et al (1998).

The superconducting gap $\Delta_s$ has been measured in the Bi-2201(La) single crystals in a wide range of temperatures $4.2$ K $\leq T \leq T_c$ by point-contact and tunnelling spectroscopy for current in **c**-direction using a conventional break-junction technique (Aminov et al 1996). After formation of a crack in the crystals at helium temperature it was easy to drive the break junctions into a point-contact regime (contact of a SNS type) with the help of a micrometer screw. In the $dI/dV$-characteristics of SNS Bi-2201(La) contacts a typical series of sharp dips at bias voltages $V_n = 2\Delta_s/en$ has been registered (Fig. 1a). The dips compose a so called subharmonic gap structure (SGS) caused by multiple Andreev reflections (Devereaux et al 1993). The highly symmetric form of the dips is consistent with a dominant s-wave pattern of the order parameter for current in **c**-direction (Devereaux et al 1993 and Klemm et al 1999).

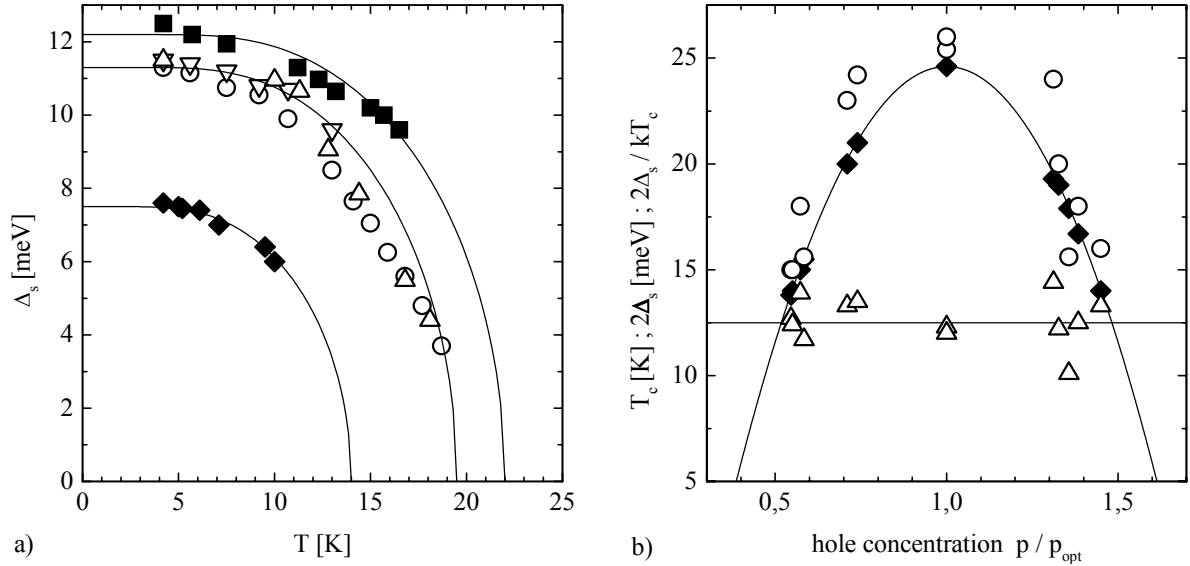

Fig. 2. a) Superconducting gap $\Delta_s$ vs temperature $T$ for three underdoped Bi-2201(La) single crystals with different $T_c$ (solid lines - BCS model).   b) Variation of $2\Delta_s$(4.2 K) (open circles), $T_c$ (solid diamonds) and $2\Delta/kT_c$ (open triangles) with the normalised hole concentration $p/p_{opt}$ for the investigated Bi-2201(La) single crystals.

We have found no noticeable difference in the form of SGS for contacts in the underdoped, optimally doped or overdoped Bi-2201(La) crystals which indicates that the symmetry of the order parameter is not changing significantly with doping. At the same time the value of the superconducting gap $\Delta_s$(4.2 K) calculated from the SGS showed scaling with $T_c$ in the whole range of doping as proposed by Deutscher (1999), with the maximum value $\Delta_{s\,max} = (12.7 \pm 0.5)$ meV corresponding to the optimal doping level ($T_{c\,max} = (24.6 \pm 0.5)$ K, $2\Delta_{s\,max}/kT_{c\,max} = 11.8 \pm 0.7$). We have obtained the same value of $\Delta_{s\,max}$(4.2 K) from the gap feature in the current-voltage characteristics (CVCs) of break junctions in the optimally doped Bi-2201(La) crystals in a tunnelling regime (contacts of SIS type). The value of $\Delta_{s\,max}$ determined in the present investigation is in reasonable agreement with $\Delta_{s\,max} = (10 \pm 2)$ meV measured by ARPES at $T = 9$ K in optimally doped Bi-2201(La) crystals (Harris et al 1997).

The most pronounced Andreev ($n = 1$) dip in the SGS ($V_1 = 2\Delta/e$) could often be observed in the $dI/dV$-characteristics even close to $T = T_c$ (Fig. 1b) which made it possible to study the temperature dependencies of the superconducting gap $\Delta_s$ for underdoped (Fig. 2a) as well as for overdoped Bi-2201(La) crystals. The main result of these studies is that the superconducting gap $\Delta_s$ definitely closes at $T = T_c$ in the whole range of doping which is again consistent with the model of Deutscher (1999).

To demonstrate the scaling of the superconducting gap $\Delta_s$(4.2 K) with the critical temperature $T_c$ for Bi-2201(La) crystals in the whole range of doping we have plotted $2\Delta_s$ and $2\Delta_s/kT_c$ vs normalised hole concentration ($p/p_{opt}$) using the empirical expression (Tallon et al 1995) $T_c/T_{c\,max} = 1 - 82.6 \cdot (p - 0.16)^2$ (Fig. 2b). It can be easily seen from Fig. 2b that within experimental errors the ratio $2\Delta_s/kT_c = 12.5 \pm 2$ does not change on doping. Furthermore this ratio exceeds noticeably the reduced-gap ratios for the optimally doped 2212- and 2223-phases ($2\Delta_s/kT_c \approx 7$) (Hudáková et al 1996). The value of $\Delta_{s\,max}$ (4.2 K) for Bi-2201(La) crystals is only two times smaller than that for the 2212-phase, so an unexpectedly low critical temperature $T_{c\,max} \approx 25$ K for the single-plane phase needs some explanation.

We have observed a new specific subharmonic gap structure due to the intrinsic multiple Andreev reflections effect (IMARE) for the current in **c**-direction. We conclude that in contrast to the 2212- and 2223-phases the $Bi_2Sr_{2-x}La_xCuO_{6+\delta}$ crystals could be treated as SNSN... superlattices (Frick et al 1992). The spacers in Bi-2201(La) crystals acquire metallic properties probably due to the presence of La. Multiple Andreev reflections of quasiparticles take place between the adjacent superconducting $CuO_2$-planes. The SGS for the stacks of $n$ Andreev contacts measured at $T < T_c$ were of extremely high quality and exactly fitted the SGS for a single SNS contact after normalisation of the voltage scale ($V \rightarrow V/n$). It was shown by Buzdin et al (1992) that in SNSN… superlattices the proximity effect can significantly increase the ratio $2\Delta/kT_c$ mainly by depressing the critical temperature $T_c$. This could possibly explain the relatively low $T_c$ in the investigated Bi-2201(La) crystals. The influence of the proximity effect on the $\Delta_s(T)$-dependencies in BSCCO samples containing intergrowths of 2212- and 2223-phases has been reported by Aminov et al (1991) and Ponomarev et al (1993) earlier.

In conclusion, we have found that the superconducting gap $\Delta_s$(4.2 K) in $Bi_2Sr_{2-x}La_xCuO_{6+\delta}$ single crystals scales with the critical temperature $T_c$ in the whole range of doping levels with the ratio $2\Delta/kT_c = 12.5 \pm 2$. The closing of the gap $\Delta_s$ at $T = T_c$ has been observed in the underdoped, optimally doped as well as in the overdoped samples. We have obtained an experimental evidence proving that the $Bi_2Sr_{2-x}La_xCuO_{6+\delta}$ crystals could be treated as superlattices of a SNSN… type, for which the ratio $2\Delta/kT_c$ is increased due to the proximity effect.

**ACKNOWLEDGMENTS**

This work was supported in part by the ISC on High Temperature Superconductivity (Russia) under the contract number 96118 (project DELTA) and by RFBR (Russia) under the contract number 96-02-18170a.

**REFERENCES**

Aminov B A et al 1991 JETP Lett. **54**, 52
Aminov B A et al 1996 Phys. Rev. B **54**, 6728
Buzdin A I, Damjanović V P and Simonov A Yu 1992 Physica C **194**, 109
Deutscher G 1999 Nature **397**, 410
Devereaux T P and Fulde P 1993 Phys. Rev. B **47**, 14638
Frick M and Schneider T 1992 Z. Phys. B - Cond. Matter **88**, 123
Harris J M et al 1997 Phys. Rev. Lett. **79**, 143
Hudáková N et al 1996 Physica B **218**, 217
Khasanova N R and Antipov E V 1995 Physica C **246**, 241
Klemm R A, Rieck C T and Scharnberg K to be published in Journ. Low Temp. Phys.
Matsuda A et al 1999 Phys. Rev. B **60**, 1377
Miyakawa N et al 1998 Phys. Rev. Lett. **80**,157
Miyakawa N et al 1999 Phys. Rev. Lett. **83**,1018
Nakano T et al 1998 J. Phys. Soc. Jpn. **67**, 2622
Ponomarev Ya G et al 1993 Journ. of Alloys & Compounds **195**, 551
Renner Ch et al 1998 Phys. Rev. Lett. **80**,149
Tallon J L et al 1995 Phys. Rev. B **51**, 12911
Vedeneev S I 1998 JETP Lett. **68**, 230
Yang W L et al 1998 Physica C **308**, 294
Yoshizaki R et al 1994 Physica C **224**, 121